\documentclass[useAMS,usenatbib]{mn2e}

\usepackage[dvips]{graphicx}
\usepackage{epsfig,natbib}
\usepackage{mathptm,amsmath,amssymb}
\usepackage{psfrag}
\usepackage{subfigure}

\def\reff@jnl#1{{\rm#1\/}}
\def\aj{\reff@jnl{AJ}}                 
\def\araa{\reff@jnl{ARA\&A}}           
\def\apj{\reff@jnl{ApJ}}               
\def\apjl{\reff@jnl{ApJ}}              
\def\apjs{\reff@jnl{ApJS}}             
\def\ao{\reff@jnl{Appl.Optics}}        
\def\apss{\reff@jnl{Ap\&SS}}           
\def\aap{\reff@jnl{A\&A}}              
\def\aapr{\reff@jnl{A\&A~Rev.}}        
\def\aaps{\reff@jnl{A\&AS}}            
\def\azh{\reff@jnl{AZh}}               
\def\baas{\reff@jnl{BAAS}}             
\def\jrasc{\reff@jnl{JRASC}}           
\def\memras{\reff@jnl{MmRAS}}          
\def\mnras{\reff@jnl{MNRAS}}           
\def\pra{\reff@jnl{Phys.Rev.A}}        
\def\prb{\reff@jnl{Phys.Rev.B}}        
\def\prc{\reff@jnl{Phys.Rev.C}}        
\def\prd{\reff@jnl{Phys.Rev.D}}        
\def\prl{\reff@jnl{Phys.Rev.Lett}}     
\def\pasp{\reff@jnl{PASP}}             
\def\pasj{\reff@jnl{PASJ}}             
\def\qjras{\reff@jnl{QJRAS}}           
\def\skytel{\reff@jnl{S\&T}}           
\def\solphys{\reff@jnl{Solar~Phys.}}   
\def\sovast{\reff@jnl{Soviet~Ast.}}    
\def\ssr{\reff@jnl{Space~Sci.Rev.}}    
\def\zap{\reff@jnl{ZAp}}               
\def\nat{\reff@jnl{Nature}}            

\def\be{\begin{equation}}
\def\ee{\end{equation}}
\def\ber{\begin{eqnarray}}
\def\eer{\end{eqnarray}}

\title[Fast cosmological parameter estimation]
{{\sc CosmoNet}: fast cosmological parameter estimation in non-flat models using neural networks}
\author[T.~Auld et al.]
{T.~Auld, M.~Bridges and M.P.~Hobson\\
Astrophysics Group, Cavendish Laboratory, Magingley Road,
Cambridge CB3 0HE, UK}

\date{Accepted ---. Received ---; in original form \today}

\pagerange{\pageref{firstpage}--\pageref{lastpage}} \pubyear{2004}

\begin{document}

\label{firstpage}

\maketitle

\begin{abstract} \noindent We present a further development of a method for accelerating the
calculation of CMB power spectra, matter power spectra and
likelihood functions for use in cosmological Bayesian inference.
The algorithm, called {\sc CosmoNet}, is based on training a
multilayer perceptron neural network.  We demonstrate the
capabilities of {\sc CosmoNet} by computing CMB power spectra (up
to $\ell=2000$) and matter transfer functions over a hypercube in
parameter space encompassing the $4\sigma$ confidence region of a
selection of CMB (WMAP + high resolution experiments) and large
scale structure surveys (2dF and SDSS). We work in the framework
of a generic 7 parameter non-flat cosmology.  Additionally we use
{\sc CosmoNet} to compute the WMAP 3-year, 2dF and SDSS
likelihoods over the same region. We find that the average error
in the power spectra is typically well below cosmic variance for
spectra, and experimental likelihoods calculated to within a
fraction of a log unit. We demonstrate that marginalised
posteriors generated with {\sc CosmoNet} spectra agree to within a
few percent of those generated by {\sc CAMB} parallelised over 4
CPUs, but are obtained 2-3 times  faster on just a \emph{single}
processor. Furthermore posteriors generated directly via {\sc
CosmoNet} likelihoods can be obtained  in less than 30 minutes on
a single processor, corresponding to a speed up of a factor of
$\sim 32$. We also demonstrate the capabilities of {\sc CosmoNet}
by extending the CMB power spectra and matter transfer function
training to a more generic 10 parameter cosmological model,
including tensor modes, a varying equation of state of dark energy
and massive neutrinos. Finally we demonstrate that using {\sc CosmoNet} likelihoods
directly, the sampling strategy adopted by {\sc CosmoMC} is highly sub-optimal. 
We find the generic {\sc Bayesys} sampler \citep{Bayesys} sampler to be a further $\sim 10$ 
times faster, 
yielding 20,000 post burn-in samples in our 7 parameter model in just 3 minutes on a single CPU.
{\sc CosmoNet} and interfaces to both {\sc
CosmoMC} and {\sc Bayesys} are publically available at {\tt
 www.mrao.cam.ac.uk/software/cosmonet}.  \end{abstract}



\section{Introduction} Bayesian inference in cosmology is normally carried out using sampling
based methods as now required by the dimensionality of the models and increasingly
high-precision of the data sets. Typically one requires the calculation of theoretical
temperature and polarisation CMB power spectra $C_\ell^{\rm TT}$, $C_\ell^{\rm TE}$,
$C_\ell^{\rm EE}$ and $C_\ell^{\rm BB}$ and/or the matter power spectrum $P(k)$ using codes
such as {\sc CMBfast} \citep{cmbfast} or {\sc CAMB} \citep{camb}. These codes typically require
of order 10 secs for spatially-flat models and 50 secs for non-flat models on a 2 GHz CPU. This
approach is therefore computationally demanding, but does have the advantage that it
is
 simple to generalise if one wishes to include new physics. {\sc CosmoMC}
\citep{cosmomc} currently represents the state of the art in cosmological Markov Chain Monte
Carlo (MCMC) sampling and employs a  number of
strategies to improve performance, such as a division of the parameter space into `slow'
parameters (which determine the evolution of structure) and `fast' parameters which determine
the primordial power spectrum. Nonetheless, the technique is still computationally expensive.

A number of examples exist in the literature of methods reliant on generating (to some degree
or other) grids of models, within which various interpolations are made to compute observable
spectra at arbitrary parameter values.  One example is {\sc Dash} \citep{Kaplinghat}
which requires a considerable investment of some 40 hours to generate a grid of transfer
functions which can then be used to generate $C_\ell$ spectra for a given parameter
combination about $30$ times faster than {\sc CAMB}.

\citet{Jimenez} have built a less demanding method around the novel idea of transformation
into the mostly uncorrelated physical  parameterisation introduced by \citet{Kosowsky}. Since
the $C_\ell$'s have a simple dependence on the input parameters they are then relatively easy
to model. The algorithm, known as {\sc CMBwarp} then uses polynomials to fit the spectra in
which the polynomial coefficients are tied to the spectra at some, single point in the
parameter space. This allows spectra to be generated $\sim 3000$ times faster than {\sc CAMB}.
Of course this method suffers from the drawback that the single model about which the
polynomial fit is specified  must be chosen carefully to lie close to the centre of the
posterior distribution as accuracy decreases away from this point. Within a $3 \sigma$ region
around the chosen model they estimate it gives better than 1\% accuracy.

The advent of larger datasets have meant the time spent calculating model likelihoods is
rapidly approaching the time necessary to generate the theoretical spectra.  {\sc CMBfit}
\citep{Sandvik} proposes to remove the step of determining spectra altogether by  providing a
semi-analytic fit directly to the WMAP likelihood as a function  of input cosmological
parameters. Given the ubiquity of WMAP data in cosmological  analyses the drawback of being
tied to a single experiment is however not as limiting as one  might think.

The methods just described, although useful, lack general applicability over a range of
theoretical spectra and datasets. We have been motivated to generate a new method that can be
applied, almost blindly to the problem of cosmological inference in order to remove the two
largest bottlenecks of theoretical spectra generation and likelihood evaluation. Previously \citet{Pico}
built a robust new method based on machine-learning called {\sc Pico}. Their method requires
the assembly of $\sim 10^4$ samples over the parameter space drawn uniformly from a desired
region that could encompass any confidence region of a given experiment. This `training set'
is  compressed via a principal component analysis (using Karhunen--Lo\`eve eigenmodes) which
typically results in a
 reduction in the dimensionality of the training set by a factor
of
 two. The training set is used to divide the parameter
 space into ($\sim 100$)
regions using $k$-means
 clustering (see e.g. MacKay 1997) with the aim of each cluster
encompassing a region of parameter space over which the power spectra vary equally. A
polynomial fit is
 then used over each cluster providing a local interpolation of the power
spectra within the cluster as a function of cosmological parameters. Crucially, the method
fails to  model the spectra accurately over the entire parameter space, hence the need for
cluster division and  thus making the algorithm difficult to extend.

Both {\sc Pico} and {\sc CMBWarp} provide similar improvements in efficiency, but {\sc
Pico} is an order of magnitude more accurate than both {\sc DASh} and {\sc CMBWarp}. It is
generic enough to be extended to any observable spectra and is flexible enough to allow
prediction of likelihood values, thus incorporating the benefits made by the {\sc CMBfit}
code. Given the current speed of the WMAP 3-year likelihood [\citet{Hinshaw}, WMAP3] code this particular facet
of the method will become extremely important in future analyses.

We previously presented \citep{Auld} a new method that combined all of the advantages of {\sc
Pico} but in a simpler and more readily expandable form by training neural networks. The
resulting algorithm is called {\sc CosmoNet} and has some considerable additional benefits in
terms of the scalability, accuracy and computational  memory requirements. In addition, the
training method we employ is sufficiently general and simple to apply that it  allows the end
user to generate their own trained nets over any chosen cosmological model.  In this paper we
extend the method to include more generic (non-flat) cosmological models, interpolations over
matter transfer functions and two large scale structure likelihoods in addition to the suite
of CMB power spectra and the WMAP3 likelihood. Additionally we extend the $\ell$ range of our
CMB spectra interpolation to $\ell_{\rm max}=2000$.
In Sec. \ref{sec:nn} we briefly describe neural
networks. In Sec. \ref{sec:results} we describe the {\sc CosmoNet} algorithm and  training
efficiency. In Sec. \ref{sec:cosmo_param} we present cosmological parameter estimates using
the trained networks implemented as {\sc CosmoNet}. In Sec. \ref{sec:10param}, we apply the
{\sc CosmoNet} training algorithm to a 10 parameter cosmological model, training the CMB power
spectra and matter transfer functions and producing parameter estimates. Our discussions and
conclusions are presented in Sec. \ref{sec:discuss}.

\section{Neural network interpolation}
\label{sec:nn} Neural networks are a methodology for computing
loosely based around the structures found in animal brains. They
consist of a number of interconnected processors called neurons.
The neurons process information separately and pass information to
one another via connections. Well-designed networks are able to
`learn' from training data and are able to make predictions when
presented with new, possible incomplete, information. For an
introduction to the science of neural networks the reader is
directed to \citet{Bailer}.

\subsection{Multilayer perceptron networks}
\label{sec:mlp}

The perceptron \citep{Rosenblatt} is the simplest type of feed-forward
neural network. It maps an input vector $\mathbf{x} \in \Re^n$ to
a scalar output $f(\mathbf{x};\mathbf{w},\theta)$ via
\begin{equation}
f(\mathbf{x};\mathbf{w},\theta) = \sum_i w_{i} x_i  +  \theta,
\label{eq:perceptron}
\end{equation}
where $\{w_{i}\}$ and $\theta$ are the parameters of the
perceptron, called the `weights' and `bias' respectively.

\begin{figure}
\begin{center}
\vspace{1cm}
\includegraphics[width=0.5\linewidth]{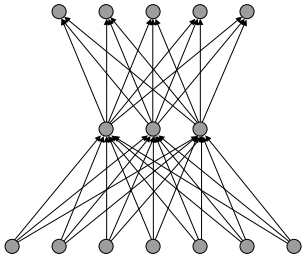}
\end{center}
\caption{\label{fig:nn} An example of a 3-layer neural network with
seven input nodes, 3 nodes in the hidden layer and five output
nodes. Each line represents one weight.}
\end{figure}

Multilayer perceptron neural networks (MLPs) are a type of
feed-forward network composed of a number of ordered layers of
perceptron neurons that pass scalar messages from one layer to the
next. In this paper, we will work with 3-layer MLPs only. They
consist of an input layer, a hidden layer and an output layer
(Fig.~\ref{fig:nn}). In such a network, the outputs of the nodes
in the hidden and output layers take the form
\begin{eqnarray}
\mbox{hidden layer:} & h_j=g^{(1)}(f_j^{(1)}); &
f_j^{(1)} = \sum_l w^{(1)}_{jl}x_l +
  \theta_j^{(1)}, \\
\mbox{output layer:} & y_i=g^{(2)}(f_i^{(2)}); & f_i^{(2)} =
\sum_j w^{(2)}_{ij}h_j + \theta_i^{(2)},
\end{eqnarray}
where the index $l$ runs over input nodes, $j$ runs over hidden nodes
and $i$ runs over output nodes. The functions $g^{(1)}$ and $g^{(2)}$
are called activation functions and are chosen to be bounded, smooth
and monotonic. In this paper, we use $g^{(1)}(x)=\tanh x$ and
$g^{(2)}(x)=x$, where the non-linear nature of the former is a key
ingredient in constructing a viable network.

The weights $\mathbf{w}$ and biases $\mathbf{\theta}$ are the
quantities we wish to determine, which we denote collectively by
$\mathbf{a}$. As these parameters vary, a very wide range of
non-linear mappings between the inputs and outputs are possible.
In fact, according to a `universal approximation theorem' (Leshno
et al. 1993), a standard multilayer feed-forward network with a
locally bounded piecewise continuous activation function can
approximate any continuous function to {\it any} degree of
accuracy if (and only if) the network's activation function is not
a polynomial. This result applies when activation functions are
chosen apriori and held fixed as $\mathbf{a}$ varies. Accuracy
increases with the number in the hidden layer and the above
theorem tells us we can always choose sufficient hidden nodes to
produce any accuracy. Since the mapping from cosmological
parameter space to the space of CMB power spectra (and WMAP3
likelihood) is known to be continuous, a 3-layer MLP with an
appropriate choice of activation function is an excellent
candidate model for the replacement of the forward model provided
by the CAMB package (and WMAP3 likelihood code).

The activation functions act as basic building blocks of
non-linearity in a neural network model and should be as simple as
possible. Additionally, the {\sc MemSys} routines used in training
(described below) require derivative information and so they
should be differentiable. The universal approximation theorem thus
motivates us to choose a monotonic (for simplicity), bounded and
differentiable function that is not a polynomial and we choose the
$\tanh$ function. Of course, this could be replaced by another such
function, such as the sigmoid function, but the interpolation
results would be almost identical.

\subsection{Network training}
\label{sec:training}

Let us consider building an empirical model of the {\sc CAMB}
mapping using a 3-layer MLP as described above (a model of the
different likelihood codes can be constructed in an analogous
manner). The number of nodes in the input layer will correspond to
the number of cosmological parameters, and the number in the
output layer will be the number of uninterpolated $C_\ell$ values
output by {\sc CAMB}. A set of training data ${\cal{D}} =
\{\mathbf{x}^{(k)},\mathbf{t}^{(k)}\}$ is provided by {\sc CAMB}
(the precise form of which is described later) and the problem now
reduces to choosing the appropriate weights and biases of the
neural network that best fit this training data.

As the {\sc CAMB} mapping is exact, this is a deterministic problem, not a
probabilistic one. We therefore wish to choose network parameters
$\mathbf{a}$ that minimise the `error' term $\chi^2(\mathbf{a})$ on the
training set given by
\begin{equation} \chi^2(\mathbf{a}) = \frac{1}{2}\sum_k
\sum_i\left[t^{(k)}_i-y_i(\mathbf{x}^{(k)};\mathbf{a})\right]^2.
\end{equation}
This is, however, a highly non-linear, multi-modal function in many
dimensions whose optimisation poses a non-trivial problem. Despite the
deterministic nature of the problem we use an extension of a Bayesian
method provided by the {\sc MemSys} package (Gull \& Skilling 1999).

The {\sc MemSys} algorithm considers the parameters $\mathbf{a}$ of the
network to be probabilistic variables with prior probability
distribution proportional to $\exp(-\alpha S(\mathbf{a}))$, where
$S(\mathbf{a})$ is the positive-negative entropy functional (\citealt{Gull}; \citealt{HobsonLasenby})
and $\alpha$ is considered a hyper-parameter of the
prior. The variable $\alpha$ sets the scale over which variations in
$\mathbf{a}$ are expected, and is chosen to maximise its marginal
posterior probability. Its value is inversely proportional to the
standard deviation of the prior.  For fixed $\alpha$, the
log-posterior is thus proportional to $- \chi^2(\mathbf{a}) + \alpha
S(\mathbf{a})$.  For each choice of $\alpha$ there is a solution
$\hat{\mathbf{a}}$ that maximises the posterior. As $\alpha$ varies, the
set of solutions $\hat{\mathbf{a}}$ is called the `maximum-entropy
trajectory'. We wish to find the maximum of $-\chi^2$ which is the
solution at the end of the trajectory where $\alpha=0$. It is
difficult to recover results for $\alpha \neq \infty$ (for large
$\alpha$ the solution is found at the maximum of the prior) when
starting with a result that lies far from the trajectory. Thus for
practical purposes, it is best to start from the point on the
trajectory at $\alpha = \infty$ and iterate $\alpha$ downwards until
either a Bayesian $\alpha$ is achieved, or in our deterministic case,
$\alpha$ is sufficiently small that the posterior is dominated by $\chi^2$.

{\sc MemSys} performs the algorithm using conjugate gradients at each step
to converge to the maximum-entropy trajectory. The required matrix of
second derivatives of $\chi^2$ is approximated using vector routines
only. This avoids the need for the $O(N^3)$ operations required to
perform exact calculations, that would be impractical for large
problems. The application of {\sc MemSys} to the problem of network training
allows for the fast efficient training of relatively large network
structures on large data sets that would otherwise be difficult to
perform in a useful time-frame.  The {\sc MemSys} algorithms are described
in greater detail in (Gull \& Skilling 1999).

\section{Results} \label{sec:results} We will demonstrate the approach of neural network
training to cosmology by attempting to replace the {\sc CAMB}
generator for the computation of CMB power spectra up to
$\ell=2000$, in both temperature and polarisation $C^{\rm
TT}_\ell$, $C^{\rm TE}_\ell$, $C^{\rm EE}_\ell$ and the matter
power spectrum $P(k)$. In general {\sc CAMB} does not compute the
CMB spectra $C_\ell$ values for all $\ell$, instead it computes
a set of 60 values (up to $\ell = 2000$) chosen at appropriately
spaced intervals to ensure coverage over the main acoustic peaks.
A cubic spline interpolation is then carried out internally in
{\sc CAMB} to produce a full compliment of $C_\ell$'s at each
$\ell$ to compare with the data. In the case of flat geometries
these chosen $\ell$ values are predetermined and fixed, but in
non-flat cases they shift, as the features of the acoustic peak
structure do with $\Omega_k$. {\sc CAMB} choses the most
appropriate $\ell$ set to ensure the main features are covered.
This creates a difficulty for our training algorithm, as one would
normally wish to learn how a set of observables changes with input
parameters. In this case the observables are actually changing. In
fact, as we shall demonstrate, if we fix the set of $\ell$'s to
those used for flat geometries, although we see some degredation
in the accuracy of the spectra we see minimal impact in the
marginalised posteriors.

In addition to the CMB power spectra, {\sc CAMB} also generates matter power spectra for
comparison with large scale structure data. We chose not to train over the spectrum directly,
but instead trained the matter transfer function $T(k)$ which can be used to generate $P(k)$
given the primordial spectrum. This has the advantage of allowing us to evaluate a number of
derived parameters such as the age of the universe and $\sigma_8$ without the need for
further trained networks \footnote{A future goal is to train networks also over the transfer functions
for CMB power spectra to achieve the same generality, but this involves substantial additional
complications and will be explored in a subsequent publication.}. Since the acoustic peak structures that appear in the CMB also
appear in the matter spectra and transfer functions, {\sc CAMB} also likes to set appropriate
scales on which to generate the spectrum in non-flat cosmologies.  In the same manner in which
we dealt with the CMB spectra we have trained the networks over a predetermined, but
sufficiently dense set of fixed $k$ values (for example {\sc CAMB} normally generates the
function at $\sim 75$ such values; in this interpolation we have used $\sim 175$). Again this approximation
has led to minimal impact on the posteriors obtained.

Current likelihood codes, such as the newly released WMAP3, now require similar computation
times to the generation of spectra. This trend is not likely to improve in the future as
larger datasets come on stream. Thus it is crucial if we are to improve the efficiency of
cosmological inference to have a combined approach for the spectra generation as well as
likelihoods. In this paper we have exploited the same network training algorithm used for
spectra to predict WMAP likelihoods as well as large scale structure likelihoods from the 2dF and SDSS
surveys. Replacement of these codes and {\sc CAMB} thus alleviates both major bottlenecks in
cosmological Bayesian inference.

\subsection{Training Data}

In order to replace the {\sc CAMB} package in codes such as {\sc CosmoMC} we need to decide
upon an appropriate region within which to train the networks. Inside this region the
regression codes reliant on the trained networks would predict the appropriate spectra and
outside this region {\sc CAMB} would need to be called in the normal fashion. Choosing too
large a region will lead to longer training periods and a reduction in the interpolation
accuracy. Too small and {\sc CAMB} would be called so often by the MCMC sampler as to render
any performance increase negligible. Training was thus carried out by uniformly sampling a $4
\sigma$ confidence region as determined  using a typical mixture of CMB and large scale
structure experiments: WMAP3 + higher resolution CMB observations (ACBAR; \citealt{ACBAR};
BOOMERang; \citealt{BOOMI}; \citealt{BOOMII}; \citealt{BOOMIII}; CBI;  \citealt{CBIII};
\citealt{CBIII} and the VSA; \citealt{Dickinson}) and galaxy surveys; 2dF; \citep{Percival} 
and SDSS; \citep{SDSSII}.

To test the approach we performed training over a non-flat cosmology parameterised by:
($\Omega_{\rm b} h^2$, $\Omega_{\rm cdm} h^2$, $\Omega_k$, $\theta$, $\tau$, $n_s$, $A_s$).
The physical parameters ($\Omega_{\rm b} h^2$, $\Omega_{\rm cdm} h^2$, $\theta$, $\tau$) were
converted back to cosmological
 parameters ($\Omega_{\rm b}$, $\Omega_{\rm cdm}$,
$H_0$,
 $z_{ \rm re}$) and used as input to CAMB to produce the training set of CMB power
spectra and matter transfer functions.  Ultimately we aim to train networks over a
sufficiently general cosmological model (see Sec. \ref{sec:10param}) so that the user could
perform any analysis over a subset of the trained parameters, setting unwanted variables to
whatever fixed value they choose. In this way the flat model computed previously in
\citet{Auld} is superceded by the results of this paper.

\subsection{Training Efficiency} To investigate training efficiency with training data set
size and number of hidden network nodes, we evaluate the testing
error as the maximum entropy trajectory is traversed.  The
training was conducted on a single $2.2$ GHz processor. Asymptotic
behaviour was observed. In particular the testing error appears to
settle down, after a period of logarithmic decrease. For a network
of this size with this amount of data it appears disproportionate
to train past $\sim 100$ hours, indeed adequate results can be
obtained in just a few hours. It is expected some tiny increase in
accuracy could be achieved for {\it much} longer training periods.
However, this would be disproportionate, unless there is a
significant error propagated through to the parameter constraints
generated by these networks.

\begin{figure}
\begin{center}
  \psfrag{ylabel}{\tiny \% Test Error}
  \psfrag{xlabel}{\tiny Time [hours]}
        \includegraphics[width=.525\columnwidth, angle = -90]{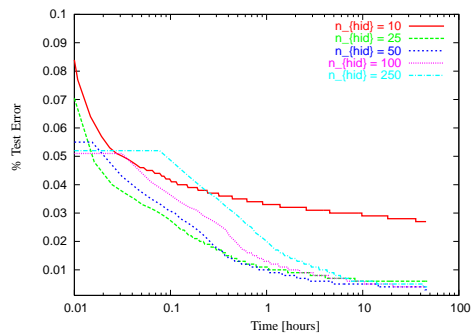}
  \caption[{{\sc CosmoNet} network selection.}]
  {\label{fig:trainingnodes} Evolution of test errors during
  training for networks with different numbers in the hidden
  layer.   Too few hidden nodes reduces accuracy, and too many slow training time.}
  \end{center}
\end{figure}

For each of the neural networks, training was then performed with 5000 training data but using
different numbers of hidden nodes. Fig. \ref{fig:trainingnodes}, shows the testing error
evolution for networks with 10, 25, 50, 100 and 250 nodes in the hidden layer, for the
$Cl^{TT}_{\ell}$ spectrum. It can be seen that increasing the numbers of nodes past 50 does
not increase accuracy, but does increase the training time. Similar experiments were then
performed to determine the optimal size of training set. Again it was observed that for each
neural network, increasing the training set size past a certain value did not increase
accuracy, but did slow training.  The optimal numbers of hidden nodes and training set sizes
obtained for all networks are displayed in Table \ref{tb:cosmonetworks}.

We note that in \citet{Habib} sub-percentage errors on the CMB spectra are
achieved for a 6 parameter flat $\Lambda$CDM model over a much larger region of
parameter space, using a Gaussian Process with just 128 training data. In this
paper we have found that of order 1000 training data produce {\it optimal} results
(for non-flat models) and we proceed on this basis. However, the reader should
note that tests showed that {\sc CosmoNet} also generated usable accuracies using 
100's rather than 1000's of training data. We do not consider the use of more
training data as a large overhead, however as the data need only be generated once, and
{\sc CosmoNet} training time scales linearly with data set size. We believe
that the method presented in \citet{Habib} would become more accurate with more
training data, but that training time may suffer, as the inversion of a matrix
is needed that requires of order the cube of the data set size operations.

\begin{table}
\begin{center}
\footnotesize
\begin{tabular}{|c|c c|}
\hline
&\hbox{Training Data} & \hbox{Hidden Nodes}\\
\hline
\hbox{CMB Spectra}  & 2000 & 50  \\
\hbox{MPT Function} & 2000 & 50   \\
\hbox{Likelihoods}  & 3000 & 50   \\
\hline
\end{tabular}
\caption{The optimal number of data, and hidden nodes in the
neural network training.} \label{tb:cosmonetworks}
\end{center}
\end{table}

\subsection{Training Results}  
\begin{figure*}
\begin{center}
  \subfigure[]{
  \psfrag{ylabel}{\tiny Error in $C_{\ell}^{TT}$}
\psfrag{xlabel}{\tiny $\ell$}
  \includegraphics[width=1.17 in, angle =-90]{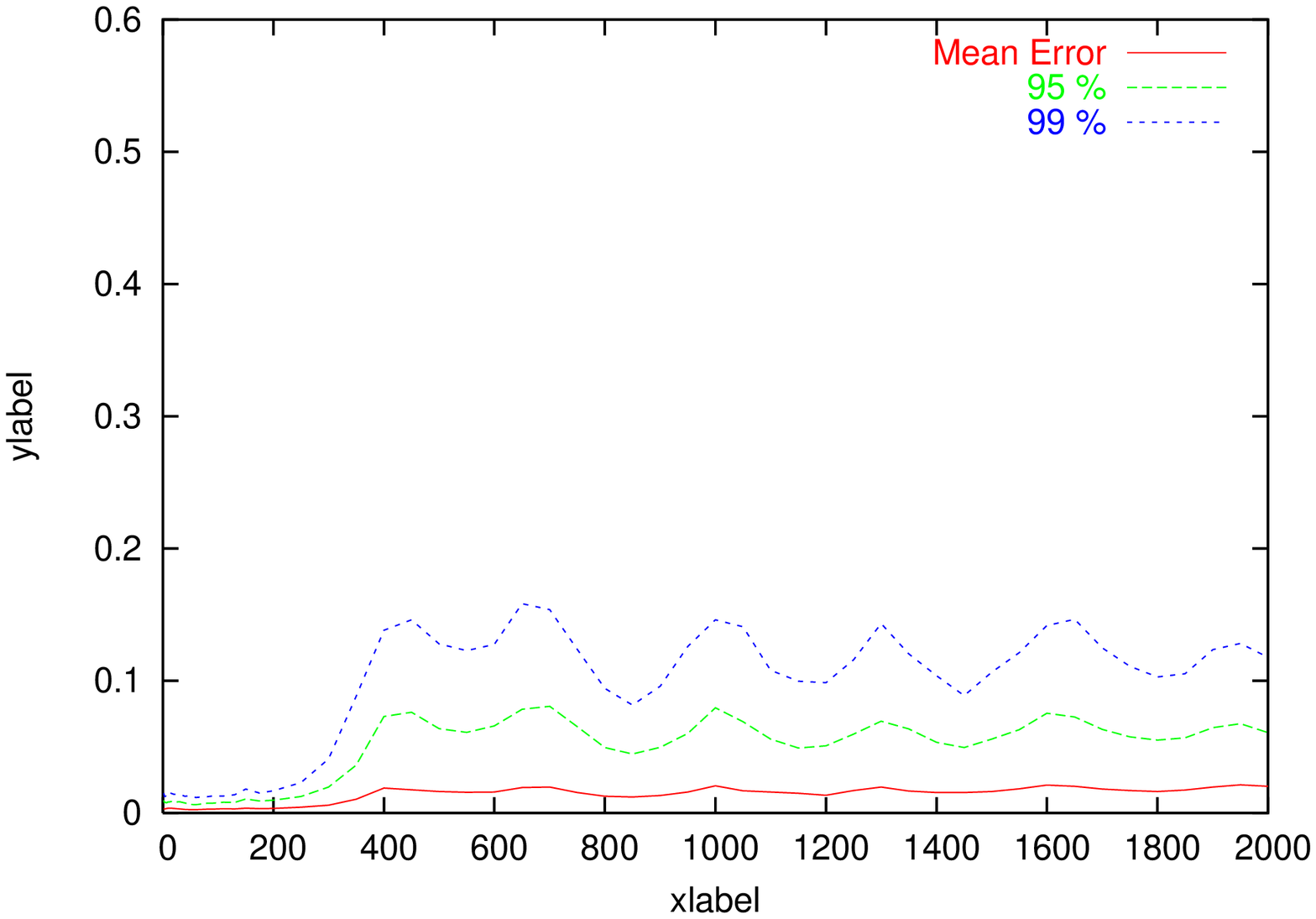}}
  \subfigure[]{
\psfrag{ylabel}{\tiny Error in $C_{\ell}^{TE}$}
  \psfrag{xlabel}{\tiny $\ell$}
  \includegraphics[width=1.17 in, angle =-90]{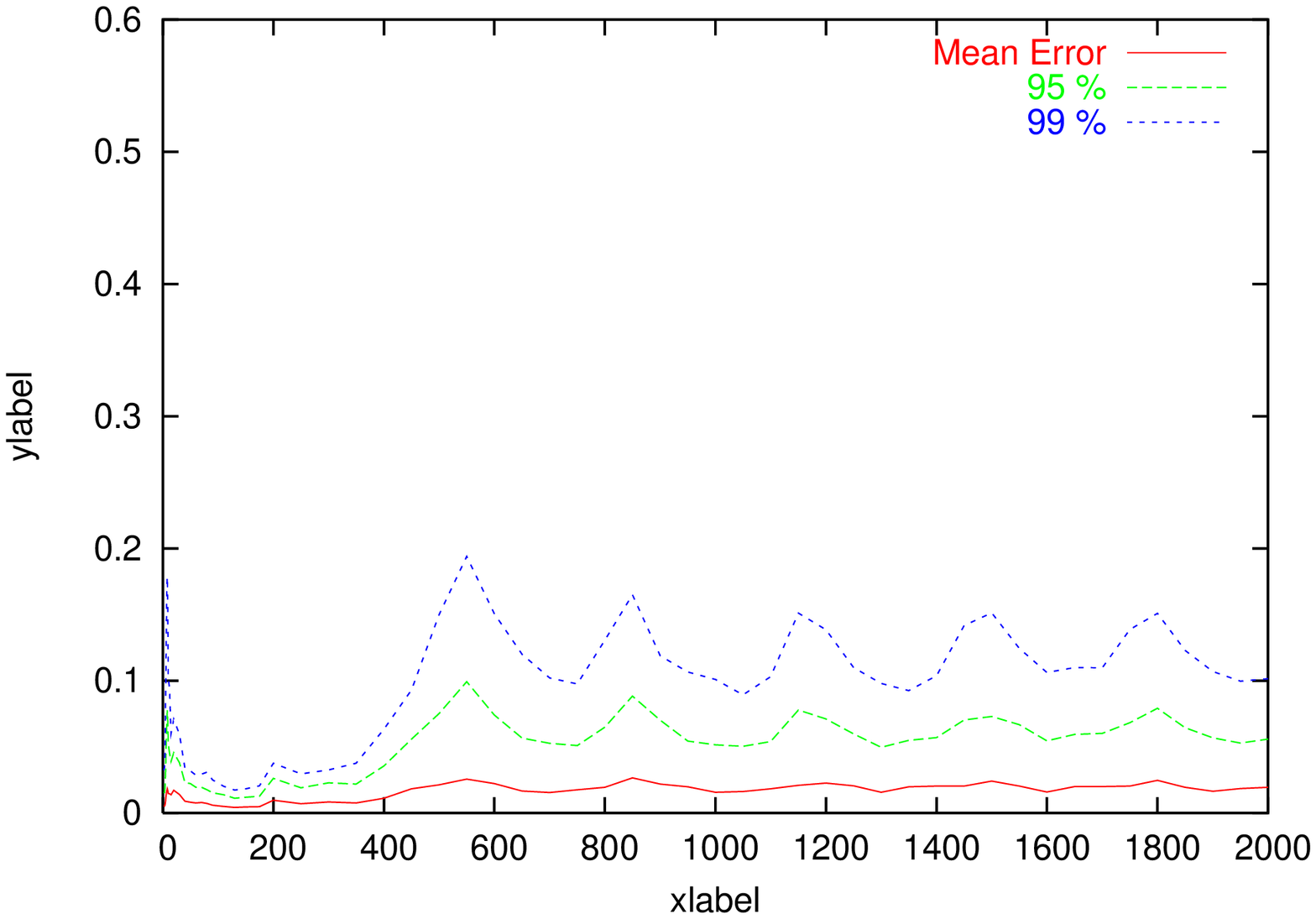}}
  \subfigure[]{
\psfrag{ylabel}{\tiny Error in $C_{\ell}^{EE}$}
  \psfrag{xlabel}{\tiny $\ell$}
  \includegraphics[width=1.17 in, angle =-90]{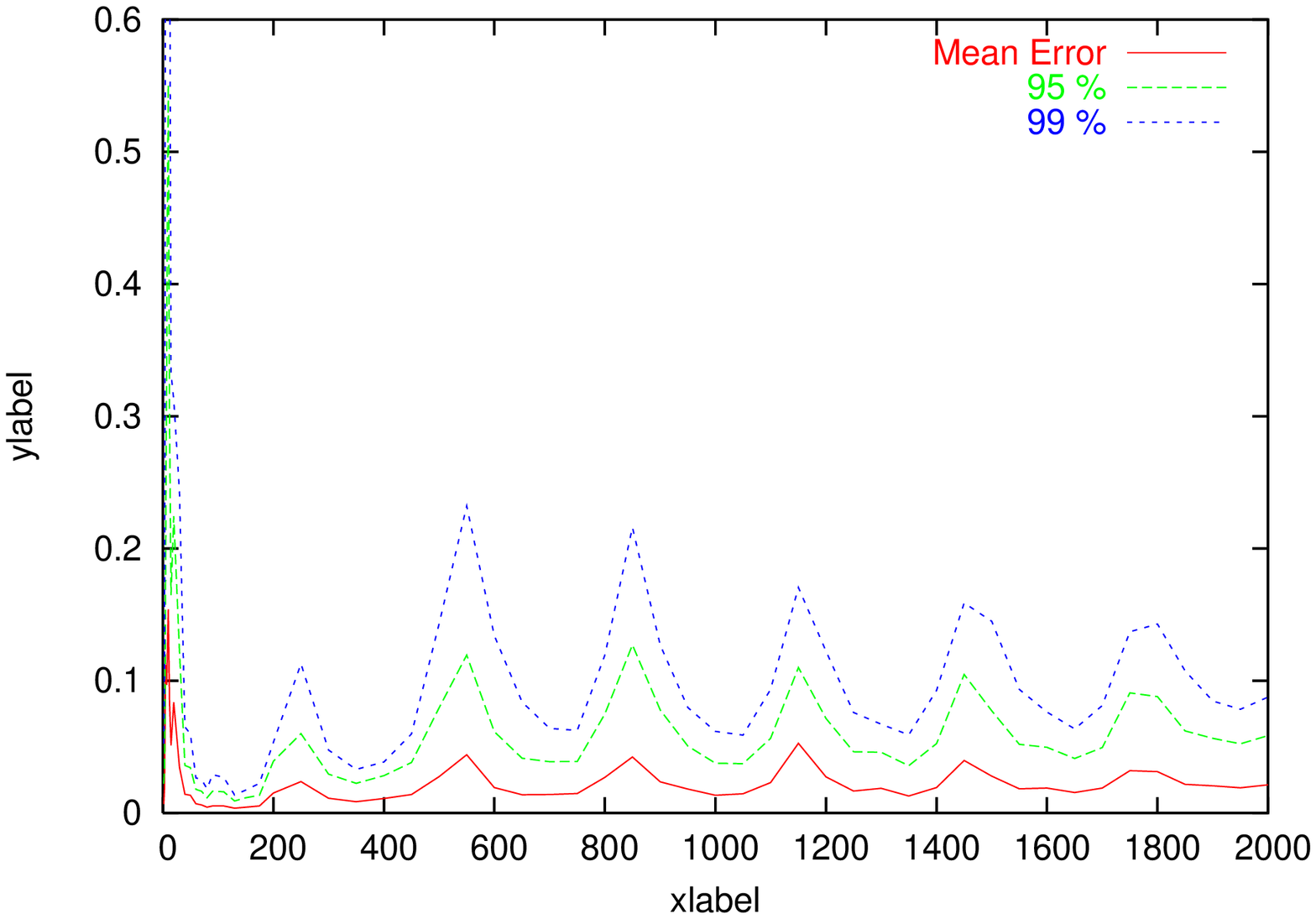}}
  \subfigure[]{
  \psfrag{ylabel}{\tiny \% Error}
\psfrag{xlabel}{\tiny k}
  \includegraphics[width=1.17 in, angle =-90]{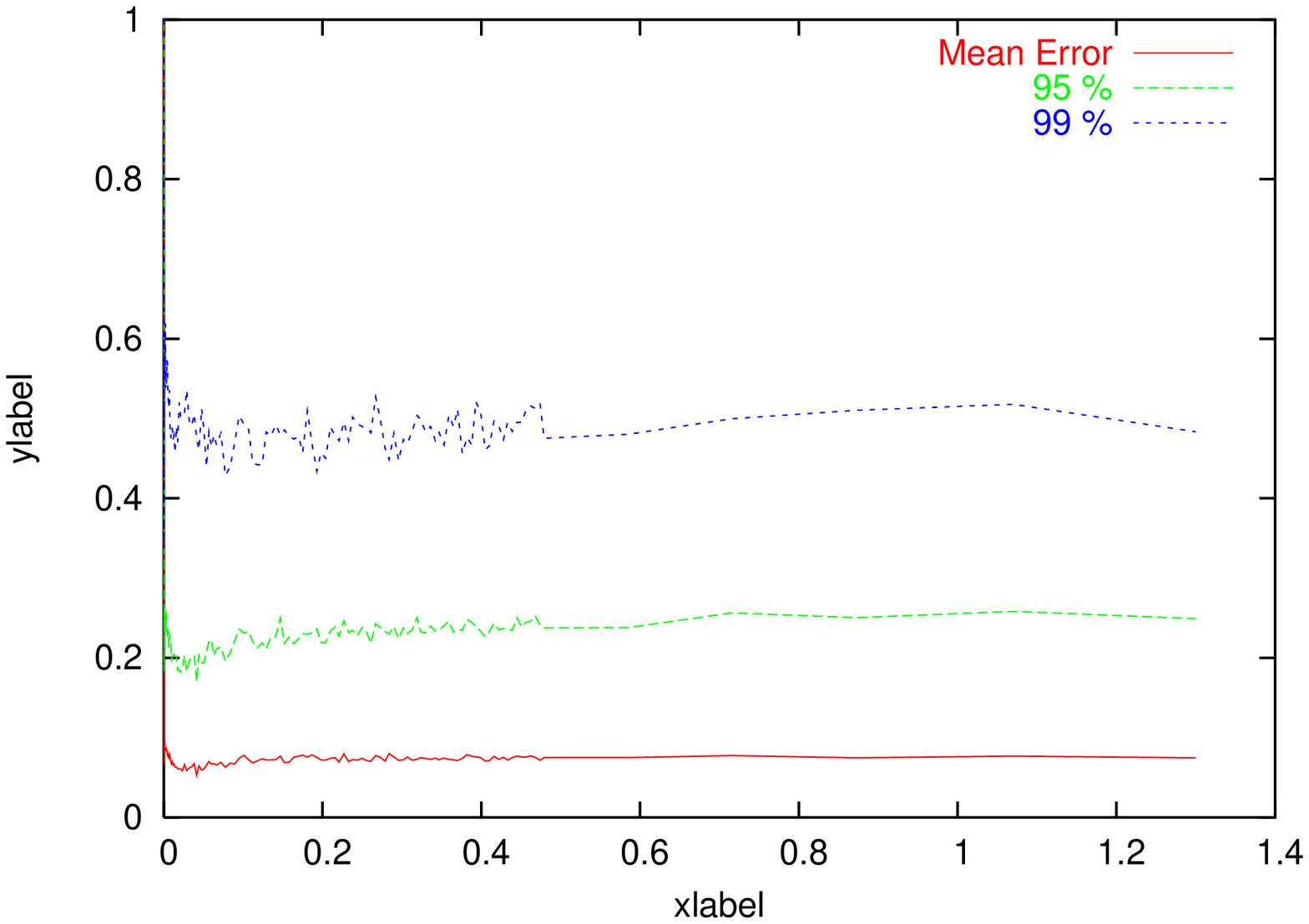}}
\caption{\label{fig:spectraB} Comparison of the performance of {\sc CosmoNet}
 versus {\sc CAMB} for TT, TE and EE power spectra (a-c) and the matter transfer function
 (d) in a 7 parameter non-flat
 cosmology. The CMB plots show the average error together with
 the 95 and 99 percentiles in units of cosmic variance. The transfer function is shown with \%
 error. }
\end{center}
\end{figure*}

Networks were trained on the optimal numbers of hidden nodes
and training data for $\sim 100$ hours. The accuracy of each
interpolated spectrum and likelihood was then evaluated on a test
set of $10^4$ models drawn uniformly from the appropriate
parameter hypercubes (see Fig. \ref{fig:spectraB}). As discussed
we would expect some error to be introduced in our interpolations
for non-flat models owing to our use of a fixed (flat) set of
$\ell$ values. We find a mean error of $\sim 5 \%$ of cosmic
variance as compared to the $\sim 1 \%$ error found in our
previous analysis of flat models \citep{Auld} which of course is
still well below any possible experimental error. More importantly
the 99 percentile errors are all comfortably below cosmic
variance, showing that the networks will be usable even when
analysing data from even a perfect experiment. A loss in accuracy
is also observed for the matter transfer interpolation.  Here we
find a mean error of less than 0.2 \%, representing a considerably
larger drop in accuracy than with the CMB spectra. However 0.2 \%
still represents a small inaccuracy given the quality of current
large scale structure datasets. The likelihood test set
correlation coefficients were all $> 0.9999$ with errors of less
than $0.2$ units close to the peak though with slightly larger
deviations away from it.

\section{Application to cosmological parameter estimation}
\label{sec:cosmo_param}

\begin{figure*}
\begin{center}
  \subfigure[]{
  \includegraphics[width=3 in]{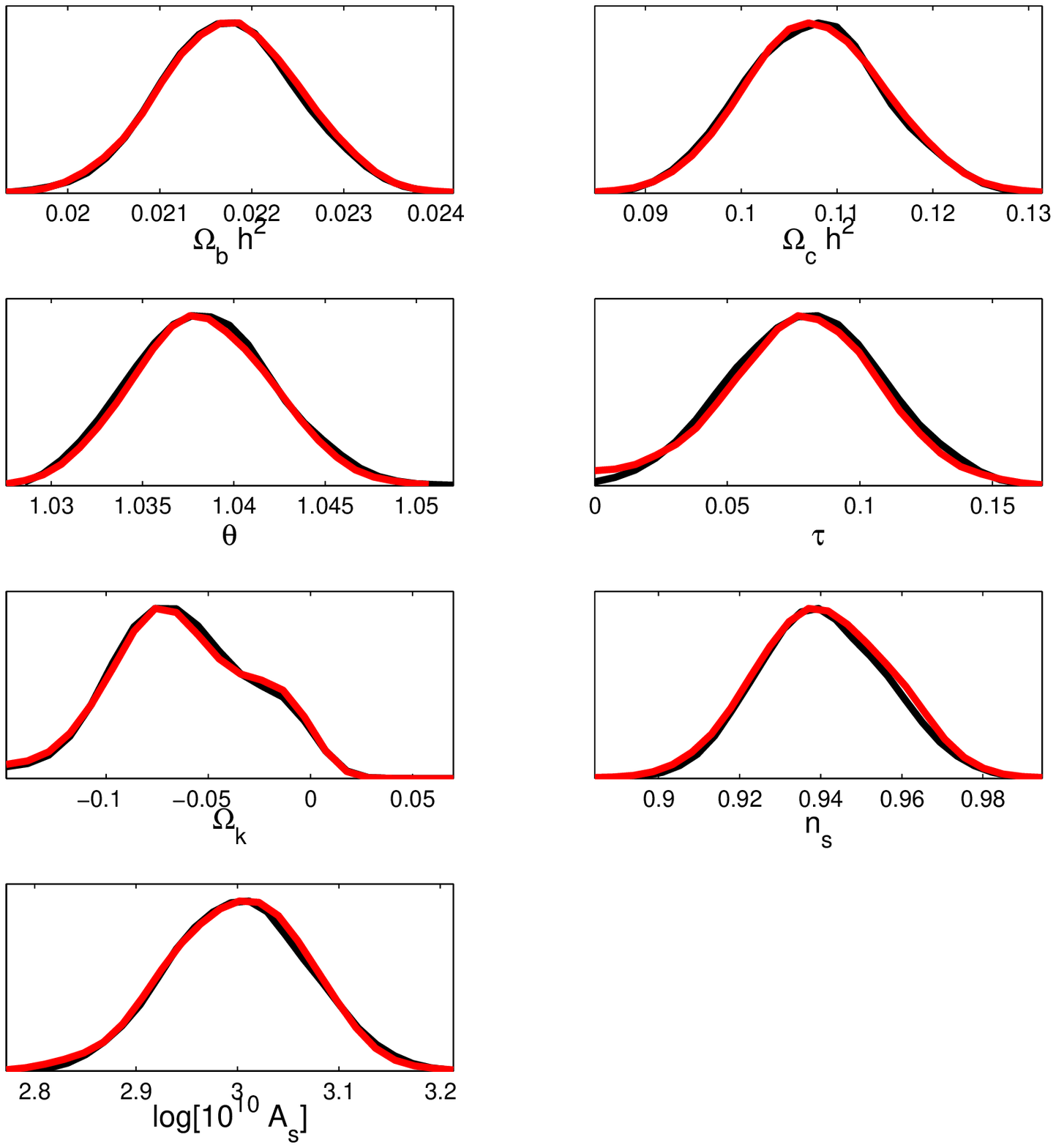}}
  \hspace{1cm}
  \subfigure[]{
    \includegraphics[width=3 in]{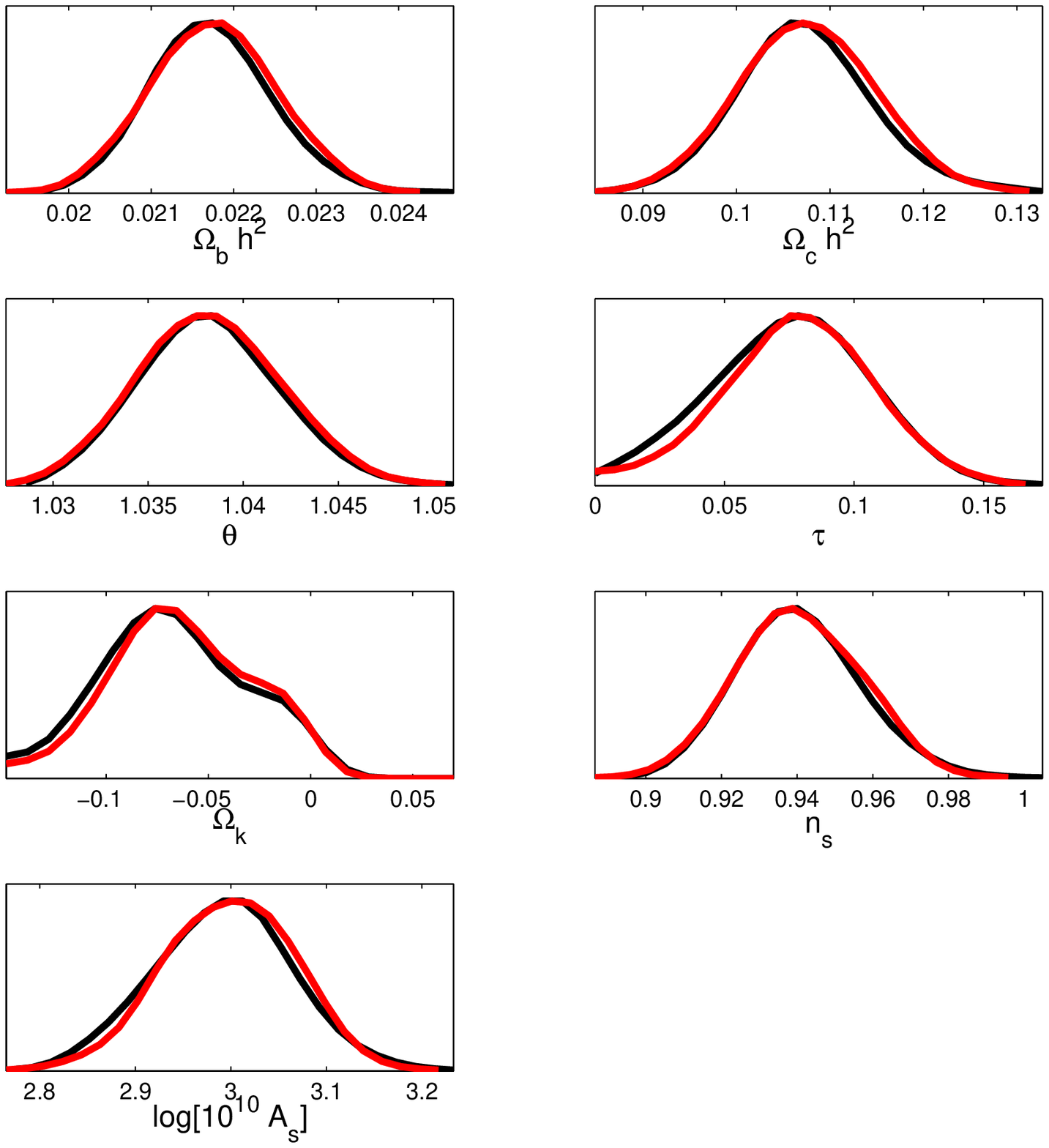}}
\caption{\label{fig:B_wlike} The one-dimensional marginalised
posteriors on the cosmological parameters within the 7-parameter non-flat
cosmology comparing: {\sc CAMB}
power-spectra and WMAP3, 2dF and SDSS likelihoods (red) with (a) {\sc CosmoNet} power
spectra and WMAP3, 2dF and SDSS likelihoods (black) and (b) {\sc CosmoNet} likelihoods (black).}
\end{center}
\end{figure*}

To illustrate the usefulness of {\sc CosmoNet} in cosmological inference we perform an
analysis of the WMAP 3-year TT, TE, EE data and 2dF and SDSS surveys using {\sc CosmoMC} in
three separate ways: (i) using {\sc CAMB} power spectra and the WMAP3, 2dF and SDSS likelihood
codes; (ii) using {\sc CosmoNet} power spectra and the WMAP3, 2dF and SDSS likelihood codes; 
(iii) using the {\sc CosmoNet} likelihood nets alone and (iv) using {\sc CosmoNet} likelihoods with
the {\sc Bayesys} sampler. The resulting marginalised parameter
constraints using each method are shown in Fig.~\ref{fig:B_wlike}, and are clearly very
similar  with mean parameter values differing by less than 1 \% of the value computed using
the standard approach (i).

To determine the speed up introduced by using {\sc CosmoNet}
spectra and likelihood interpolations, 4 parallel MCMC chains were
run on Intel Itanium 2 processors at the COSMOS cluster (SGI Altix
3700) at DAMTP, Cambridge using the basic {\sc CosmoMC} sampling
package. The time required to generate $\sim 20000$ post burn-in
MCMC samples was recorded using methods (i)--(iv) described above
\footnote{Note that {\sc CAMB} was in fact parallelised over 3
additional processors per chain, therefore totalling 16 CPUs}. The
results (see Table \ref{table:timings}) illustrate that using {\sc
CosmoNet} spectra one can obtain reasonable posterior
distributions in roughly 8 hours on a \emph{single} CPU per chain
whereas using {\sc CAMB} not only took between 2-3 times longer
but required 3 additional CPUs per chain. Using {\sc CosmoNet}s
likelihood interpolations alone produced dramatic time savings,
with accurate results in roughly 30 minutes.

Cosmologists have invested considerable time in developing samplers that have as efficient a
proposal distribution as possible. In {\sc
 CosmoMC}, the multi-variate Gaussian proposal
distribution has a covariance matrix that is regularly updated using statistics from
 the samples
gathered up to that point. This does lead to a higher acceptance rate and a corresponding lower
number of likelihood
 evaluations, but is computationally intensive in its own right.
 However,
when using a {\sc CosmoNet} likelihoods directly there is no need to reduce the number of likelihood
calls. The process of updating the proposal distribution slows the task considerably, as can
be seen when comparing times with the efficient, yet likelihood intensive {\sc Bayesys}
sampler \citep{Bayesys} via method (iv), computing the relevent posteriors in just 3 minutes. 

The reader should also note from both our previous work \citep{Auld},
and that of the 10 parameter models below, that the
timings for parameter estimation are roughly
independent of the number of model parameters used. Our
regression algorithm is indifferent to the complexity of the input
cosmology.

\begin{table}
\begin{center}
\footnotesize
\begin{tabular}{|c||c||c||c||c|}
\hline
\hbox{\bf Method} & \hbox{(i)} & \hbox{(ii)} & \hbox{(iii)} & \hbox{(iv)}\\
\hbox{No. chains} & 4 & 4 & 4 & 4\\
\hbox{No. CPU/chain} & 4 & 1 & 1 & 1\\
\hbox{Run time} & $>$ 16 hrs. & $\sim 8$ hrs. & $\sim 30$ mins. & $\sim 3$ mins.\\
\hline
\end{tabular}
\caption{Time required to gather $\sim$ 20,000 post burn-in MCMC samples using different combinations
of {\sc
 CAMB}, {\sc CosmoNet}, the experimental likelihood codes and {\sc Bayesys}. Note that {\sc CAMB} is
parallelised in method
 (i) over 4 CPUs per chain, if a single processor were used these timings
would approach 4 $\times$ that quoted.}
\label{table:timings}
\end{center}
\end{table}

\section{Towards a 10 dimensional parameter space} \label{sec:10param}  In \citet{Auld} we
presented trained networks capable of replacing {\sc CAMB} and experimental likelihood codes
for a 6 parameter flat cosmology. In this paper we have shown that this method is easily
extendable to the more arduous computational demands of a non-flat cosmology. To test the
scaleability to even higher dimensions we now examine a 10 dimensional cosmology including, in
addition to the basic 7 given in Sec. \ref{sec:results}: the equation of state of
dark energy, $w$, the neutrino mass fraction, $\nu$ and the tensor to scalar ratio, $r$.

\begin{figure}
\begin{center}
\psfrag{xlabel}{\tiny $\ell$}
  \subfigure[]{
  \psfrag{ylabel}{\tiny Error in $C_{\ell}^{TT}$}
  \includegraphics[width=1.13 in, angle =-90]{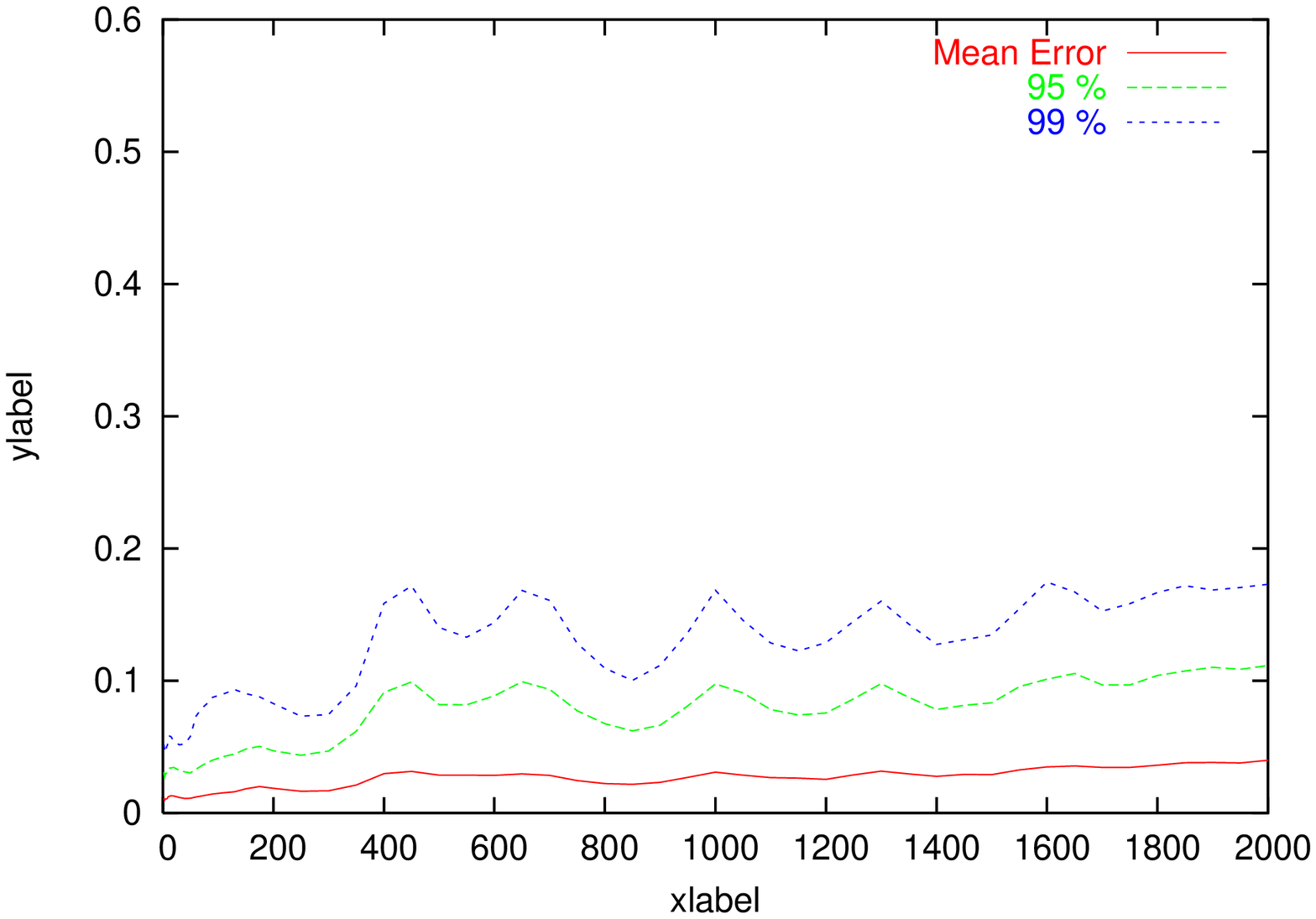}}
  \subfigure[]{
\psfrag{ylabel}{\tiny Error in $C_{\ell}^{TE}$}
  \includegraphics[width=1.13 in, angle =-90]{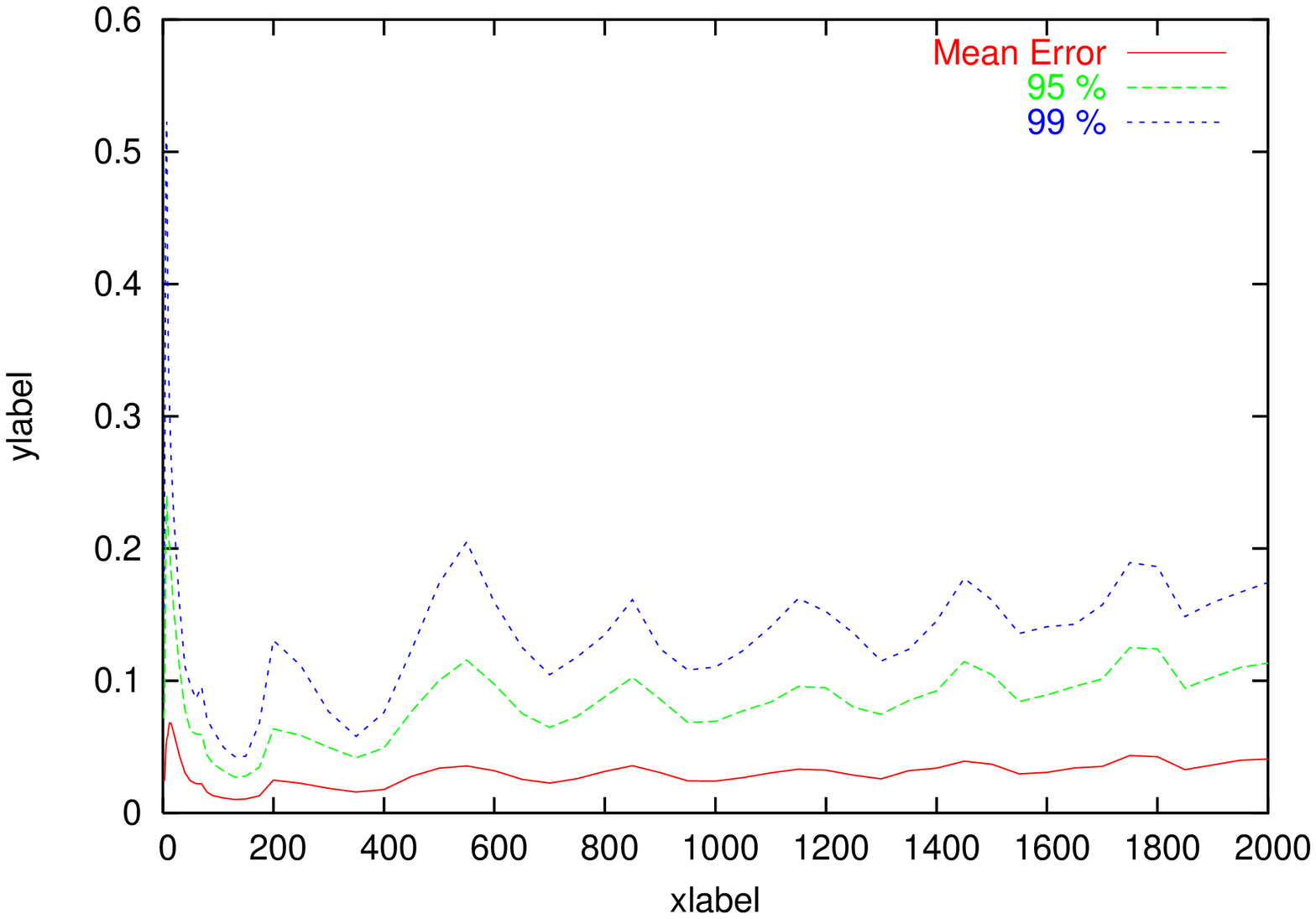}}\\
\subfigure[]{
\psfrag{ylabel}{\tiny Error in $C_{\ell}^{EE}$}
  \includegraphics[width=1.13 in, angle =-90]{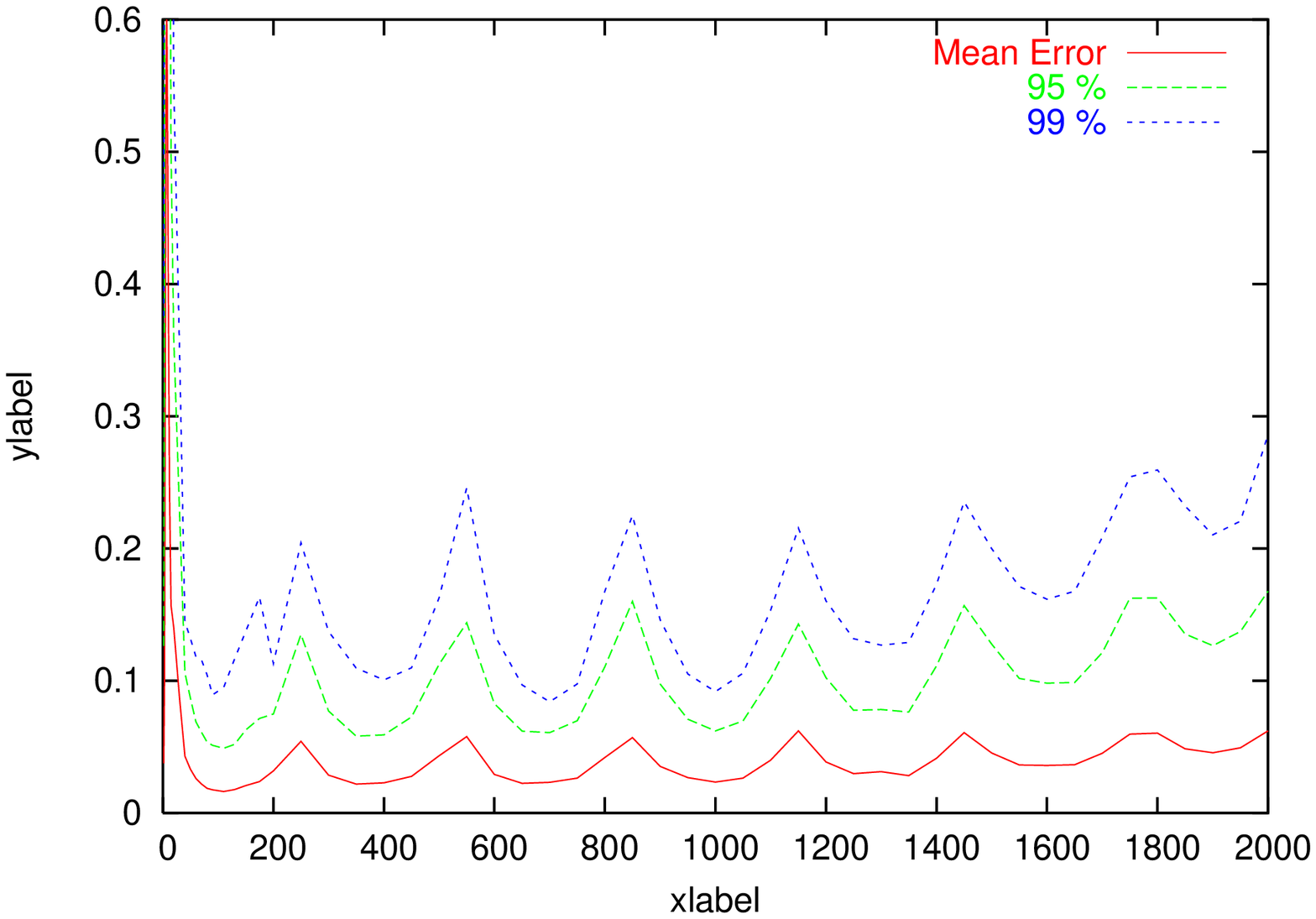}}
\subfigure[]{
\psfrag{ylabel}{\tiny Error in $C_{\ell}^{BB}$}
  \includegraphics[width=1.13 in, angle =-90]{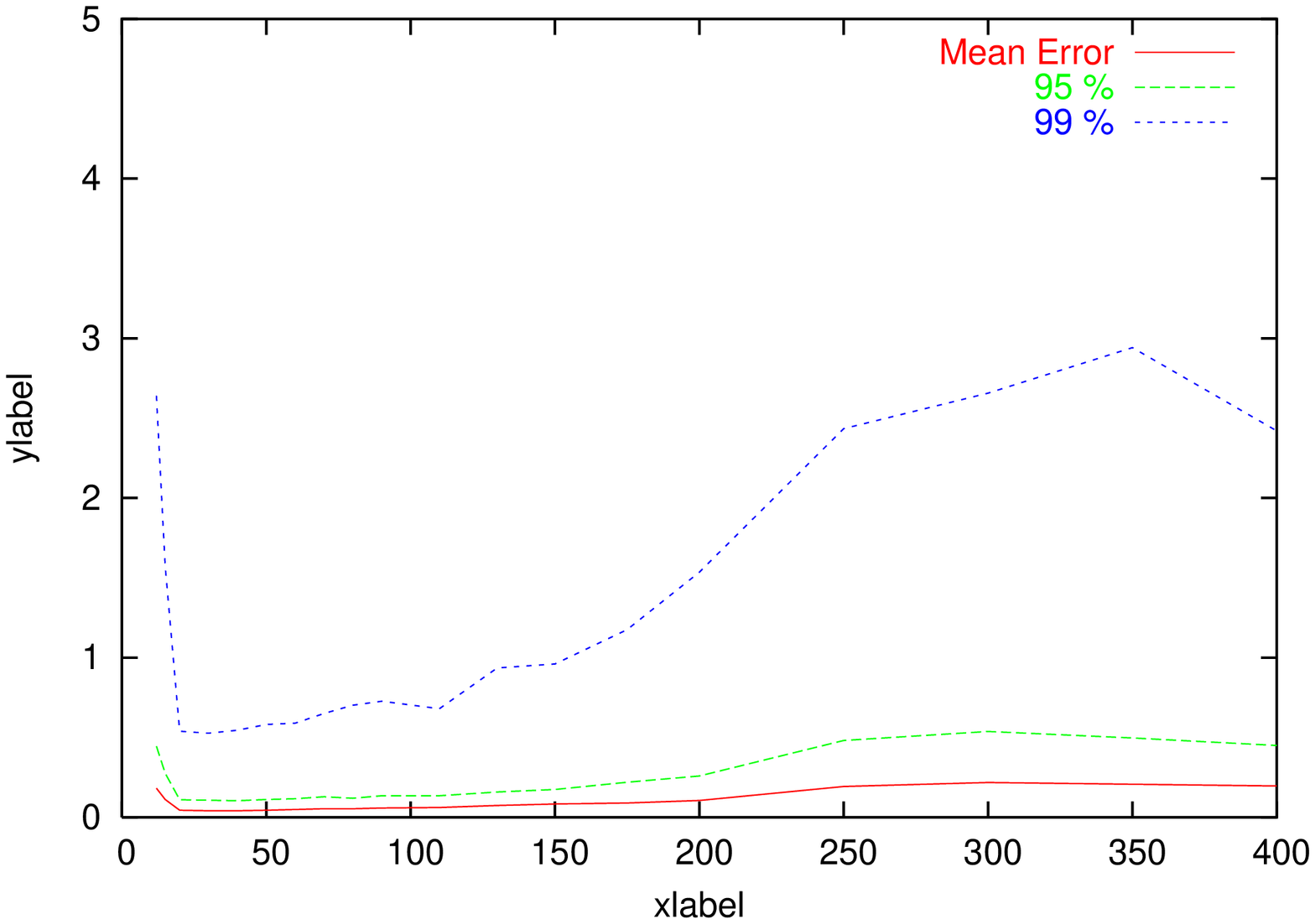}}\\
  \subfigure[]{
  \psfrag{ylabel}{\tiny \% Error}
\psfrag{xlabel}{\tiny k}
\includegraphics[width=1.13 in, angle =-90]{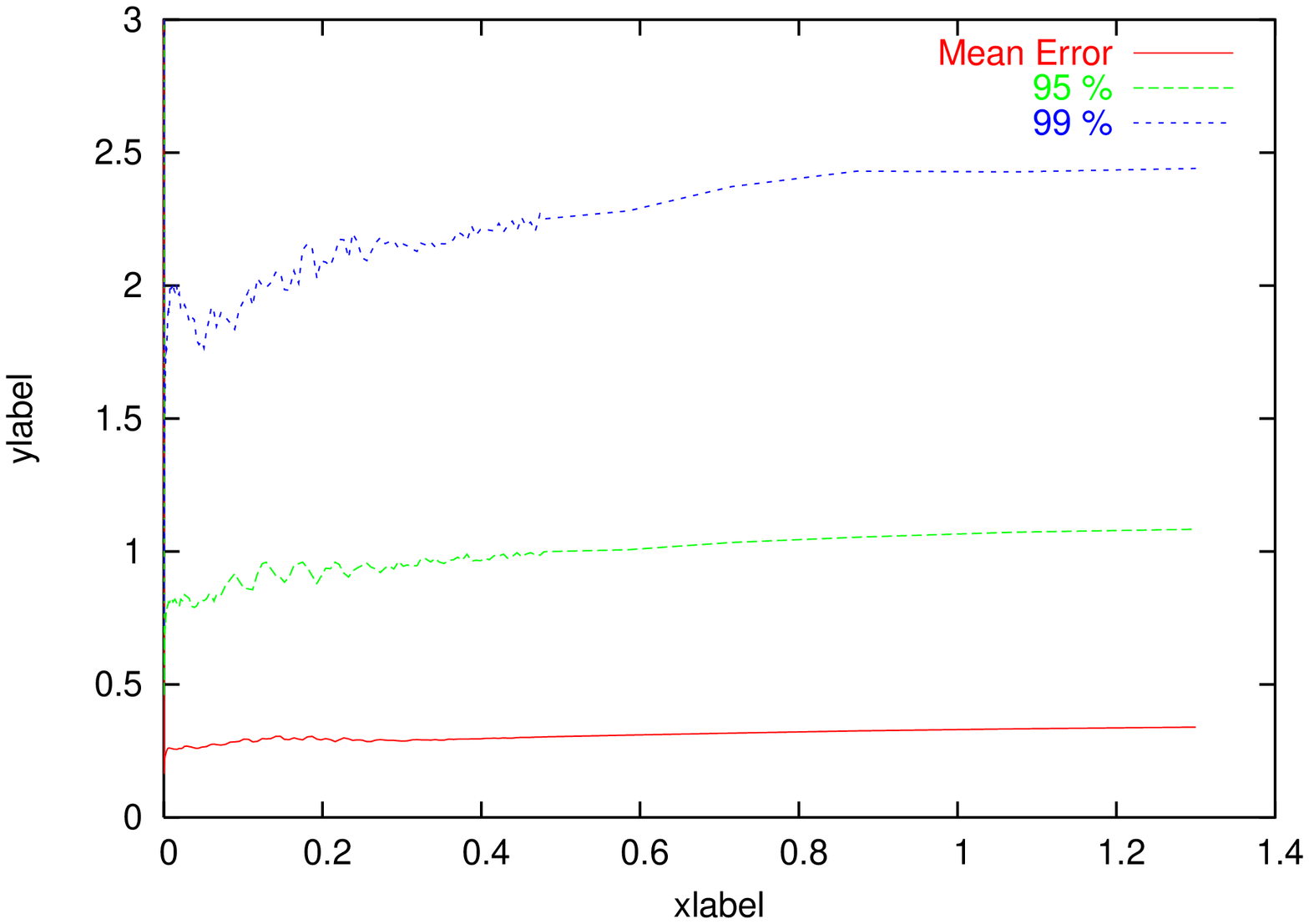}}
\caption{\label{fig:spectraC} Comparison of the performance of {\sc CosmoNet}
 versus {\sc CAMB} for TT, TE, EE and BB power spectra (a-d) and matter transfer function (d)
 in a 10 parameter non-flat
 cosmological model. The CMB plots show the average error together with
 the 95 and 99 percentiles in units of cosmic variance. The transfer function is shown with \%
 error. }
\end{center}
\end{figure}

Training efficiency was examined as per the 7 parameter model (see
Table \ref{table:training_10param}) and it was found that little
increase in the quantity of training data or training time was
needed for optimum results for the CMB power spectra and matter
transfer function. An accurate tracer of the scaling of the
training algorithm is given by the number of network hidden nodes
as this determines the amount of computational resource required.
In this case we find that at worst a $50 \%$ increase in the
number of hidden nodes is needed  for a $\sim 40 \%$ rise in the
number of parameter dimensions (going from 7 to 10). This
represents slightly more than a linear rise in resources and
demonstrates our algorithm is easily scaleable to even higher
dimensions if necessary. The accuracy of interpolated CMB spectra
and matter transfer functions did not decrease \emph{at all} when
compared to the 7 parameter interpolations (see Fig.
\ref{fig:spectraC}), suggesting that the largest source of error
in our method is introduced by fixing the set of $\ell$ and $k$
values at their flat positions. Providing an accurate
interpolation for the three likelihood surfaces was however
problematic. Using the order of 1000 training data provides very
sparse coverage of the 10 dimensional hypercube. For CMB power
spectra and the matter transfer function this is not a problem
since they vary smoothly over a limited dynamical range. For
likelihoods however, the dynamical range is much larger and to
obtain parameter constraints we need very good accuracy within a
region having a volume of the order of that of a $1 \sigma$
hypersphere. This hypersphere has a volume over 400,000 times less
than the 10 parameter $4 \sigma$ hypercube over which we performed
the training. This suggests that much more training data would be
required. A potential solution to this problem would be to train
likelihood networks only in some region over which the likelihood
value was within (say) 50 log units of the peak value. The shape
of this region could be determined by a classification net that
returns an output that predicts whether a point lies inside or
outside the desired region. This method will be explored in a
future publication.

\begin{table}
\begin{center}
\footnotesize
\begin{tabular}{|c|c c|}
\hline
&\hbox{Training Data} & \hbox{Hidden Nodes}\\
\hline
\hbox{CMB Spectra}  & 2000 & 75  \\
\hbox{MPT Function} & 2000 & 50   \\
\hline
\end{tabular}
\caption{The required number of data, and hidden nodes in
the neural network training for optimum performance in a 10 dimensional parameterisation.}
\label{table:training_10param}
\end{center}
\end{table}

\begin{figure}
\begin{center}
  \includegraphics[width=3 in]{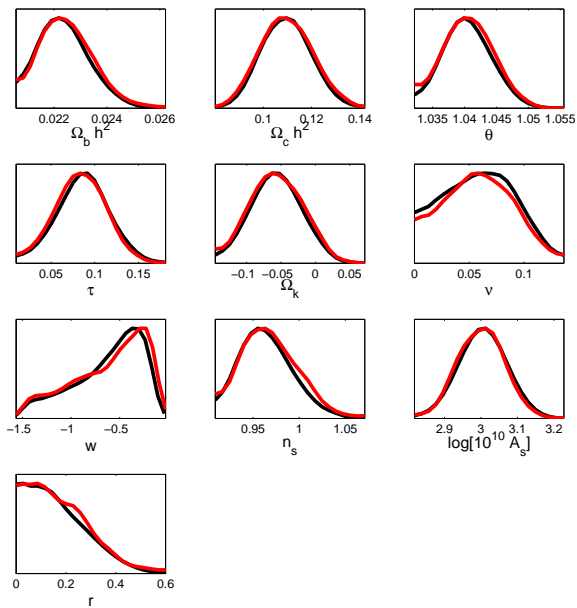}
\caption{\label{fig:C} The one-dimensional marginalised
posteriors on the cosmological parameters within the 10-parameter non-flat
$\Lambda$CDM model including tensor modes, varying equation of state of dark energy and massive
neutrinos comparing: {\sc CAMB}
power-spectra and WMAP3, 2dF and SDSS likelihoods (red) with {\sc CosmoNet} power
spectra and WMAP3, 2dF and SDSS likelihoods (black).}
\end{center}
\end{figure}

Marginalised posteriors obtained from {\sc CosmoNet} spectra were
found to be accurate to within a few $\%$ of those computed via
{\sc CAMB} (see Fig. \ref{fig:C}), and took roughly 8 hours on a
single CPUs per chain to calculate (see Table
\ref{table:timings_10param}). {\sc CAMB} however required more
than 20 hours of computational time with parallelisation over a
further 4 CPUs per chain.

\begin{table}
\begin{center}
\footnotesize
\begin{tabular}{|c||c||c|}
\hline
\hbox{\bf Method} & \hbox{(i)} & \hbox{(ii)}\\
\hbox{No. chains} & 4 & 4 \\
\hbox{No. CPU/chain} & 4 & 1 \\
\hbox{Run time} & $>$ 20 hours & $\sim 8$ hours\\
\hline
\end{tabular}
\caption{Time required to gather $\sim$ 20,000 post burn-in MCMC samples using different
combinations
of {\sc CAMB}, {\sc CosmoNet} and the experimental likelihood codes. Note that {\sc CAMB}
is parallelised in method (i) over 4 CPUs per chain, if a single processor were used these  timings
would approach 4 $\times$ that quoted.}
\label{table:timings_10param}
\end{center}
\end{table}

\section{Discussion and Conclusions} \label{sec:discuss}  We have extended our method of
accelerating the estimation of CMB and matter power transfer functions, WMAP, 2dF and SDSS
likelihood evaluations based on training a multilayer perceptron neural network to more
generic non-flat cosmologies. We have demonstrated that the use of trained neural networks
such as {\sc CosmoNet} can replace the bulk of computational effort required by cosmological
evolution codes such as {\sc CAMB} and experimental likelihood codes, like that of WMAP3.
{\sc CosmoNet} shares all the improvements made by {\sc Pico} in terms of accuracy on both
spectral interpolation and parameter constraints, but has now been scaled to a more generic 7
parameter non-flat cosmology. Furthermore, although the training procedure requires the
optimisation of a highly non-linear multi-dimensional  function, the end user simply runs the
{\sc MemSys} package essentially as a `black box'. This means {\sc CosmoNet} remains simple
and efficient to train. We have found the biggest bottleneck in the procedure to be the
generation of training and testing data using {\sc CAMB}. Increasing the model complexity had
limited impact on the necessary training time (all models taking about 100 hours to train) or
interpolation accuracy. Moreover the increase in network hidden nodes was at worst linear with
increasing parameter space. Thus we expect few resource difficulties in extending this method
to even higher dimensions.

Although accurate likelihood interpolations in the 10 dimensional
model interpolation are currently beyond the reach of our method,
the corresponding CMB spectra and matter transfer functions
\emph{are} sufficiently accurate allowing a speed up over the
standard performance of {\sc CosmoMC}. 

Finally, replacing the {\sc CosmoMC} sampler entirely with {\sc Bayesys} can produce  further
dramatic time savings of a factor of $\sim 10$, computing $\sim 20,000$ post burn-in samples in a
few minutes on a single CPU. 

\subsection*{ACKNOWLEDGMENTS}
TA acknowledges a studentship from EPSRC. MB was supported by a Benefactors' Scholarship
at St. John's College, Cambridge and an Isaac Newton Studentship. This work was
conducted in cooperation with SGI/Intel utilising the Altix 3700 supercomputer
at DAMTP Cambridge supported by HEFCE and PPARC. We thank S. Rankin and V.
Treviso for their assistance.

\bibliographystyle{mn2e}
\bibliography{netpaperII}

\begin{thebibliography}{}

\bibitem[\protect\citeauthoryear{{Auld}, {Bridges}, {Hobson} \& {Gull}}{{Auld}
  et~al.}{2007}]{Auld}
{Auld} T.,  {Bridges} M.,  {Hobson} M.~P.,    {Gull} S.~F.,  2007, \mnras, 376,
  L11

\bibitem[\protect\citeauthoryear{{Bailer-Jones}}{{Bailer-Jones}}{2001}]{Bailer}
{Bailer-Jones} C.,  2001, Automated Data Analysis in Astronomy.
Narosa Publishing House, New Delhi

\bibitem[\protect\citeauthoryear{{Dickinson} et~al.,}{{Dickinson}
  et~al.}{2004}]{Dickinson}
{Dickinson} C.,  et~al., 2004, \mnras, 353, 732

\bibitem[\protect\citeauthoryear{{Fendt} \& {Wandelt}}{{Fendt} \&
  {Wandelt}}{2007}]{Pico}
{Fendt} W.~A.,  {Wandelt} B.~D.,  2007, \apj, 654, 2

\bibitem[\protect\citeauthoryear{{Gull} \& {Skilling}}{{Gull} \&
  {Skilling}}{1999}]{Gull}
{Gull} S.,  {Skilling} J.,  1999, Quantified maximum entropy: MemSys 5 users'
  manual.
Maximum Entropy Data Consultants Ltd, Royston

\bibitem[\protect\citeauthoryear{{Habib}, {Heitmann}, {Higdon}, {Nakhleh} \&
  {Williams}}{{Habib} et~al.}{2007}]{Habib}
{Habib} S.,  {Heitmann} K.,  {Higdon} D.,  {Nakhleh} C.,    {Williams} B.,
  2007, ArXiv Astrophysics e-prints

\bibitem[\protect\citeauthoryear{{Hinshaw} et~al.,}{{Hinshaw}
  et~al.}{2006}]{Hinshaw}
{Hinshaw} G.,  et~al., 2006, ArXiv Astrophysics e-prints

\bibitem[\protect\citeauthoryear{{Hobson} \& {Lasenby}}{{Hobson} \&
  {Lasenby}}{1998}]{HobsonLasenby}
{Hobson} M.~P.,  {Lasenby} A.~N.,  1998, \mnras, 298, 905

\bibitem[\protect\citeauthoryear{{Jimenez}, {Verde}, {Peiris} \&
  {Kosowsky}}{{Jimenez} et~al.}{2004}]{Jimenez}
{Jimenez} R.,  {Verde} L.,  {Peiris} H.,    {Kosowsky} A.,  2004, \prd, 70,
  023005

\bibitem[\protect\citeauthoryear{Jones et~al.,}{Jones  et~al.}{2006}]{BOOMII}
Jones W.~C.,  et~al., 2006, \apj, 647, 823

\bibitem[\protect\citeauthoryear{{Kaplinghat}, {Knox} \&
  {Skordis}}{{Kaplinghat} et~al.}{2002}]{Kaplinghat}
{Kaplinghat} M.,  {Knox} L.,    {Skordis} C.,  2002, \apj, 578, 665

\bibitem[\protect\citeauthoryear{{Kosowsky}, {Milosavljevic} \&
  {Jimenez}}{{Kosowsky} et~al.}{2002}]{Kosowsky}
{Kosowsky} A.,  {Milosavljevic} M.,    {Jimenez} R.,  2002, \prd, 66, 063007

\bibitem[\protect\citeauthoryear{Kuo et~al.,}{Kuo  et~al.}{2004}]{ACBAR}
Kuo C.-l.,  et~al., 2004, Astrophys. J., 600, 32

\bibitem[\protect\citeauthoryear{{Lewis} \& {Bridle}}{{Lewis} \&
  {Bridle}}{2002}]{cosmomc}
{Lewis} A.,  {Bridle} S.,  2002, \prd, 66, 103511

\bibitem[\protect\citeauthoryear{{Lewis}, {Challinor} \& {Lasenby}}{{Lewis}
  et~al.}{2000}]{camb}
{Lewis} A.,  {Challinor} A.,    {Lasenby} A.,  2000, \apj, 538, 473

\bibitem[\protect\citeauthoryear{Montroy et~al.,}{Montroy
  et~al.}{2006}]{BOOMIII}
Montroy T.~E.,  et~al., 2006, \apj, 647, 813

\bibitem[\protect\citeauthoryear{{Percival} et~al.,}{{Percival}
  et~al.}{2001}]{Percival}
{Percival} W.~J.,  et~al., 2001, \mnras, 327, 1297

\bibitem[\protect\citeauthoryear{Piacentini et~al.,}{Piacentini
  et~al.}{2006}]{BOOMI}
Piacentini F.,  et~al., 2006, \apj, 647, 833

\bibitem[\protect\citeauthoryear{Readhead et~al.,}{Readhead
  et~al.}{2004}]{CBIII}
Readhead A. C.~S.,  et~al., 2004, Astrophys. J., 609, 498

\bibitem[\protect\citeauthoryear{{Rosenblatt}}{{Rosenblatt}}{1958}]{Rosenblatt}
{Rosenblatt} F.,  1958, Psychological Review, 65, 386

\bibitem[\protect\citeauthoryear{{Sandvik}, {Tegmark}, {Wang} \&
  {Zaldarriaga}}{{Sandvik} et~al.}{2004}]{Sandvik}
{Sandvik} H.~B.,  {Tegmark} M.,  {Wang} X.,    {Zaldarriaga} M.,  2004, \prd,
  69, 063005

\bibitem[\protect\citeauthoryear{{Seljak} \& {Zaldarriaga}}{{Seljak} \&
  {Zaldarriaga}}{1996}]{cmbfast}
{Seljak} U.,  {Zaldarriaga} M.,  1996, \apj, 469, 437

\bibitem[\protect\citeauthoryear{Skilling}{Skilling}{2004}]{Bayesys}
Skilling J., , 2004, BayeSys3 Users Manual

\bibitem[\protect\citeauthoryear{Tegmark et~al.,}{Tegmark
  et~al.}{2004}]{SDSSII}
Tegmark M.,  et~al., 2004, Astrophys. J., 606, 702

\end{thebibliography}

\bsp 
\label{lastpage}
\end{document}